\documentclass[aps,prb,twocolumn,superscriptaddress,showpacs]{revtex4-1}
\usepackage{graphicx}
\usepackage{amssymb}
\usepackage{amsmath}
\usepackage{appendix}
\usepackage{comment}

\usepackage{soul}

\usepackage{tikz}       
\usepackage{tikz-feynman}
\makeatletter
\tikzfeynmanset{compat=\tikzfeynman@version@major.\tikzfeynman@version@minor.\tikzfeynman@version@patch}
\makeatother

\begin{document}

\title{Quantum-kinetic theory of spin-transfer torque and magnon-assisted transport in nanostructures}

\author{Scott A. Bender}
\affiliation{Institute for Theoretical Physics, Utrecht University, Leuvenlaan 4, 3584 CE Utrecht, The Netherlands}
\author{Rembert A. Duine}
\affiliation{Institute for Theoretical Physics, Utrecht University, Leuvenlaan 4, 3584 CE Utrecht, The Netherlands}
\affiliation{Department of Applied Physics, Eindhoven University of Technology, PO Box 513, 5600 MB Eindhoven, The Netherlands}
\author{Yaroslav Tserkovnyak}
\affiliation{Department of Physics and Astronomy, University of California, Los Angeles, California 90095, USA}

\begin{abstract}
We theoretically investigate the role of spin fluctuations in charge transport through a magnetic junction. Motivated by recent experiments that measure a nonlinear dependence of the current on electrical bias, we develop a systematic understanding of the interplay of charge and spin dynamics in nanoscale magnetic junctions. Our model captures two distinct features arising from these fluctuations: magnon-assisted transport and the effect of spin-transfer torque on the magnetoconductance. The latter stems from magnetic misalignment in the junction induced by spin-current fluctuations. As the temperature is lowered, inelastic quantum scattering takes over thermal fluctuations, exhibiting signatures that make it readily distinguishable from magnon-assisted transport.

\end{abstract}

\maketitle

\section{Introduction}
\label{intro}

The accurate electrical detection and control of the spin degree of freedom of electrons remains a central goal of spintronics. An early success of the field was the demonstration of magnetoresistance in conducting magnetic multilayers, allowing for the determination of the magnetic state of a heterostructure via its electrical resistance.\cite{Binasch:1989he,Baibich:1988ib,Levy:1990ej} Later shown was the possibility of writing magnetic states by the application of large current densities. \cite{Mangin:2006ix,Hayakawa:2005ic,Huai:2004bd,Kubota:2005gx,Albert:2000gv,Kubota:2007cd,Stiles:2002dz} An electrical current traversing the structure becomes spin polarized by one magnetic layer and exerts a spin-transfer torque (STT) on another \cite{Slonczewski:1996vc,Berger:1996is,Zhang:2002df}. If the STT is large enough to overcome damping, it can induce switching between magnetic states with different electrical resistance, thereby paving the way for a current-driven ``write" complement to the ``read" functionality of magnetoresistance. \cite{Hosomi:2005jv,Diao:2006gn}

In structures with large macrospins, it typically suffices to characterize charge transport in linear response.   However, as devices are scaled down and the spins of the components become smaller, magnetic fluctuations become increasingly important and can give rise to new interplays between magnetic dynamics and electrical transport, which may manifest through nonlinear charge transport features. 

One known example of this interplay is magnon-assisted transport (MAT) originating from inelastic electron-magnon scattering. \cite{Zhang:1997kx} A second effect stems from changes in the magnetoresistance of a heterostructure caused by STT-altered misalignments of the magnetic components. \cite{Kozub:2007gv} While such misalignments arise from thermal fluctuations at high temperatures, recent work has argued that in nanostructures at low temperatures, such an effect is quantum mechanical in nature. \cite{Zholud:2017ed}

As both MAT and STT may manifest as zero-bias kinks in the electrical response, a careful theoretical treatment is necessary to distinguish these, as well as to elucidate the nature (classical versus quantum) of the STT in nanoscale junctions.\cite{Zhang:2017bs} Previous theoretical studies, such as quantum master equation approaches \cite{Wang:2012cb,*Wang:2013eq}, treat quantum spin fluctuations, but focus on spin dynamics rather than charge transport features. Others, including a quantum Green's function approach \cite{Mahfouzi:2014hu}, formally integrate spin fluctuations and focus on the ensuing magnetotransport.

In this paper, we develop a self-consistent treatment for spin fluctuations coupled to the electrical response of a magnetic heterostructure, incorporating the relevant inelastic processes and including both thermal and quantum fluctuations on equal footing. Developing a quantum rate equation for magnetic dynamics, in particular, allows for the phenomenological inclusion of dissipation to the environment and the incorporation of pertinent nonlinearities.   Our model allows for a parsing and comparison of the contributions of the different effects to overall electrical response. Our approach yields simple, analytic equations that, in addition to laying bare the underlying physics, makes clear the temperature, bias and junction-size regimes in which different effects dominate. 

Our model offers two key insights. (1) While our results are compatible with the interpretation of [\onlinecite{Zholud:2017ed}], we go beyond a phenomenological description of magnon emission. The low-temperature STT predicted by our model, which includes quantum fluctuations, results in a zero-bias resistance kink.  Our model also describes the crossover from classical (thermal) STT at high temperatures to the quantum behavior at low temperatures, which is marked by a change in the resistance from a monotonic dependence on bias to a local extremum. (2) We show that both types of STT may be readily distinguished from MAT by reversing the relative magnetic orientation in the junction, as well as by the bias scales on which they appear.  Additionally, we predict that both MAT and STT give rise to a flat resistance at biases smaller than the magnetic field energy, which may be observed at higher fields.

The paper is organized as follows. In Sec.~\ref{hamiltonian}, we start by briefly discussing the physical processes behind MAT and STT, after which we introduce our model Hamiltonian for spin-dependent electron transport, which we generalize to include spin fluctuations. In Sec.~\ref{charge_trans}, using the Kubo formula, we then compute the charge transport resulting from this generalized Hamiltonian, including both inelastic and elastic electron hopping.  The inelastic current is MAT, while the elastic current includes magnetoresistance, which depends on the magnetic state. Next, in Sec~\ref{scsd} we determine the steady-state magnetic dynamics by computing the spin transfer rate (i.e., the STT) from the same inelastic scattering processes; the resulting expression for the magnetic steady-state is finally reinserted back into the current to give the full, self-consistent charge dynamics. Finally, on the basis of the resulting expression for the current, in Sec~\ref{elec_respons} we discuss the contributions of both STT and MAT to the nonlinear $I-V$ curves over different ranges of temperature and bias. 

\section{Model}
\label{hamiltonian}

\begin{figure}[pt]
\includegraphics[width=\linewidth,clip=]{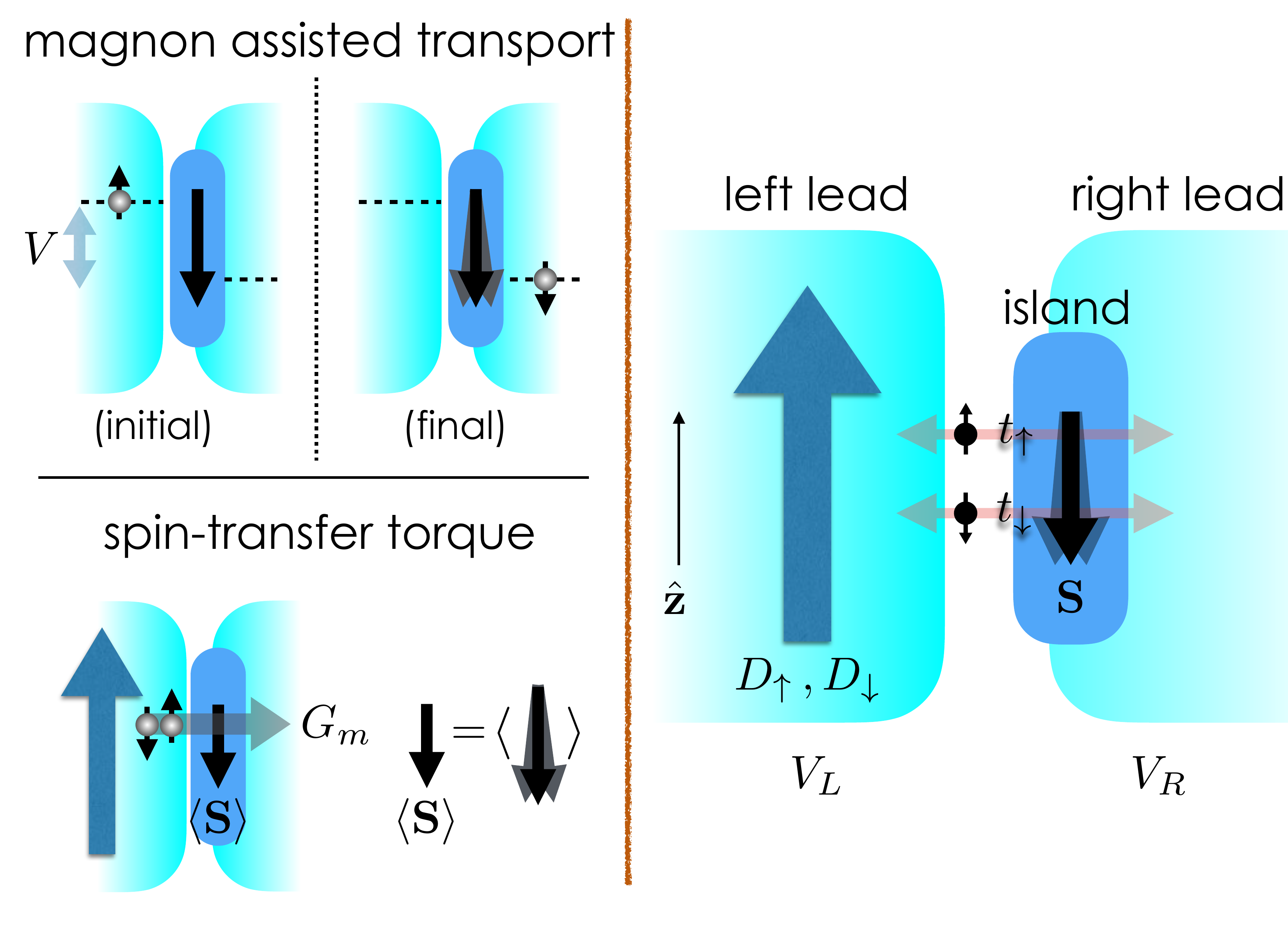}
\caption{Upper left figure: magnon-assisted transport (MAT). Electrons excite or destroy magnons as the move across a heterostructure. Lower left figure: spin-transfer torque (STT) alteration of the magnetoresistance. The STT alters fluctuations of the spin $\mathbf{S}$, affecting the $\langle \mathbf{S}\rangle$ and thereby the magneto conductance $G_m$. Right figure: MTJ schematic. The magnetic island does not hold charge, but the island spin $\mathbf{S}$, which is allowed to fluctuate, gives rise to spin-dependent hopping between the left and right-leads.}.
\label{sch}
\end{figure} 

Before introducing our model, we briefly discuss the physics of MAT and STT. Consider a conducting magnetic junction, in contact with metallic reservoirs. Under an electrical bias, electrons flow from one energy reservoir, across the junction, to the other.  If the bias is large enough, some of the electrons may undergo spin-flip scattering, creating a magnetic excitation, a magnon, of the magnetic components of the junction. The electron correspondingly loses some energy to a magnon, entering a lower energy state in the other energy reservoir (see upper left of Fig.~\ref{sch}). This inelastic transport, i.e. MAT, has been shown to manifest as a zero-bias kink in the electrical response.  \cite{Zhang:1997kx,Han:2001er,Lu:2003hv,Balashov:2008ih} 

Meanwhile, the junction magnetoconductance $G_m$ depends generally on the relative orientations of magnetic layers; because spin-fluctuations of the different layers change, on average, the relative orientations of the layers, the magnetoresistance obtains a correction $\delta G_m$ proportional to the amplitudes of the fluctuations. Under a bias, the STT may enhance or reduce the amplitude of the spin-fluctuations, \cite{Tsoi:2000jh,Katine:2000eb,Kiselev:2003hy,Madami:2011cx} altering $\delta G_m$ and self-consistently changing the current $I$ flowing across the heterostructure.  As a result, the current may demonstrate a nonlinear dependence on bias. At high temperatures, these spin-fluctuations are thermal in nature (coresponding to \textit{classical} STT). However, the dependence of the magnetoconductance on bias has been observed as a zero-bias kink in the differential resistance of a nanopillar spin valve, which is  known to persist at low temperatures.\cite{Sun:2008ik,Manchon:2009do} In [\onlinecite{Zholud:2017ed}], it was argued that such features may arise from \textit{quantum} STT. Importantly, this STT incorporates spontaneous magnon emission, which gives rise to a dependence of the resistance on the absolute value $|I|$ of the current, thereby manifesting as the zero-bias kink: remarkably, spin fluctuations are enhanced for both directions of current, which stands in contrast to classical predictions. \cite{Zhang:2017bs}

We now turn to the task of writing a minimal model that captures both effects. To incorporate inelastic spin-flip scattering of electrons naturally, we focus on a magnetic tunnel junction (MTJ) (oriented in a parallel or antiparallel configuration). While some quantitative differences in charge transport may arise between sequential tunneling and coherent electron transport through a metallic devices such as that in Ref.~[\onlinecite{Zholud:2017ed}], we expect our results to be qualitatively generic (barring special features associated with mesoscopic resonances, Coulomb blockade, or any band-structure anomalies).

In our model the role of magnetization of one of the leads is ascribed to the spin of a magnetic nano-island connecting the leads (see right of Fig.~\ref{sch}), which gives rise to spin-dependent charge transport through the junction barrier. The island spin is allowed to fluctuate, coupling magnetic fluctuations with charge transport. Charge transport occurs between a metallic lead on the right and a magnetic metallic lead on the left, between which electrons map hop. We suppose that both leads are large reservoirs whose properties are unaffected by transport, with the magnetic polarization of the left-lead fixed in the $+z$ direction; in addition we suppose that the spin of the left-lead is sufficiently large that its fluctuations may be neglected. While the right-lead is nonmagnetic, we consider the spins of electrons traversing the junction to be coupled to the spin $\mathbf{S}$ of the magnetic island, which connects the leads but does not hold electrons; the amplitudes for electron hopping from one lead to the other thus depend on the orientation of the electron spin relative to $\mathbf{S}$.  For a \textit{fixed}, semiclassical $\mathbf{S}$, the corresponding tunneling Hamiltonian can be written generally:
\begin{equation}
\hat{\mathcal{H}}_T=\sum_{\sigma\sigma'}\sum_{\nu\nu'}\gamma_{\nu\nu'}^{(\sigma\sigma')} \hat{b}_{\nu\sigma}^\dagger \hat{a}_{\nu'\sigma'}+H.c.\, ,
\label{genT}
\end{equation}
where the amplitudes $\gamma_{\nu\nu'}^{(\sigma\sigma')}$ depend on the unit vector $\mathbf{n}=\mathbf{S}/S\hbar$ of the island spin, with $S=|\mathbf{S}|/\hbar$ as the island spin. Here $\hat{a}_{\nu'\sigma'}$ and $\hat{b}_{\nu\sigma}$ are annihilation operators for electrons in left-lead eigenstate $\nu'$ and spin $\sigma'$ and right-lead eigenstate $\nu$ and spin $\sigma$, respectively; the indices $\sigma,\sigma'=\uparrow,\downarrow$ denote spin orientations relative to the $z$ direction. Note that we do not include magnetism of the left-lead in the tunneling Hamiltonian, Eq.~(\ref{genT}), but will subsequently include it through the left-lead density of states. For a model isotropic in spin space, we may expand in powers of $\mathbf{n}\cdot \check{\boldsymbol{\sigma}}$, where $\check{\boldsymbol{\sigma}}$ is a vector of Pauli matrices, with $\check{\dots}$ denoting $2\times 2$ spin-structure. One obtains the semiclassical expression:
\begin{equation}
\check{\gamma}_{\nu\nu'}=A_{\nu\nu'}\check{I}+B_{\nu\nu'} \mathbf{n}\cdot \check{\boldsymbol{\sigma}}\, ,
\label{gamma}
\end{equation}
(with $\check{I}$ as the 2$\times$2 identity) which is general in the absence of magnetism in the leads and constitutes an isotropic Kondo model. 

To parametrize thermal and quantum fluctuations of the tunnel island macrospin, we quantize $\mathbf{S}$ via the Holstein-Primakoff transformation: 
\begin{equation}
\hat{S}_{z}\equiv \hat{c}^{\dagger}\hat{c}-S,\, \,\, \hat{S}_{-}\equiv \hat{S}_{x}-i\hat{S}_{y}=\sqrt{2S-\hat{c}^{\dagger}\hat{c}}\hat{c}\approx\sqrt{2S}\hat{c}\, ,
\label{hp}
\end{equation}
where $\hat{c}$ is a bosonic magnon annihilation operator.  We will restrict ourselves to biases below the STT-induced classical instability,\cite{Berger:1996is} so that the average direction of $\mathbf{n}$ is fixed and fluctuations are incoherent: $\langle \hat{c} \rangle=0$.  In writing Eq.~(\ref{hp}), we have chosen the direction of $\langle \mathbf{S} \rangle$ to be oriented antiparallel (i.e. in the $-z$ direction) to the left-lead spin in equilibrium; thus a magnon, which carries spin opposite to the spin order parameter, is associated with a quantum $+\hbar \mathbf{z}$ of angular momentum.  (The other case we will consider, the parallel orientation, may be obtained by changing the sign of the left reservoir polarization $P_L$). For simplicity, we specialize to fluctuations of the macrospin only, with higher energy magnon modes assumed to be energetically inaccessible. In addition, we restrict ourselves to a large spin $S\gg 1$ and small magnon occupation numbers, $N= \langle \delta \hat{S}_z \rangle=\langle \hat{c}^\dagger \hat{c}\rangle  \ll S$, allowing for the expansion of the radical in Eq.~(\ref{hp}). 

 Writing $\mathbf{n}$ as $\mathbf{S}/S\hbar$ in Eq.~(\ref{gamma}) and inserting Eq.~(\ref{hp}), we find the tunneling Hamiltonian, Eq.~(\ref{genT}), has two physically distinct contributions: $\mathcal{H}_T\approx \mathcal{H}_e+\mathcal{H}_i$. The first term, 
\begin{equation}
\mathcal{H}_e=\sum_{\sigma =\pm }\sum_{\nu,\nu' }\hat{t}_{\nu\nu'}^{(\sigma)}\hat{b}^\dagger_{\nu \sigma} \hat{a}_{\nu' \sigma}+H.c.\, ,
\end{equation}
conserves magnon number and gives rise to elastic scattering of electrons through the tunnel junction. Here, $\hat{t}_{\nu \nu'}^{(\pm)}=A_{\nu\nu'}\mp B_{\nu\nu'}+2B_{\nu\nu'}\hat{c}^\dagger\hat{c}/S$  (with $\pm$ denoting spin orientation in the positive (negative) $\mathbf{z}$ direction) capture mixing of the spin-dependent hopping amplitudes by fluctuations of the island spin. The second term, 
\begin{align}
\mathcal{H}_i=\frac{1}{\sqrt{S/2}}\sum_{\nu\nu'}B_{\nu\nu'}\left(\hat{c}^{\dagger}b_{\nu\downarrow}^{\dagger}a_{\nu'\uparrow}+\hat{c}b_{\nu \uparrow}^{\dagger}a_{\nu '\downarrow}\right)+H.c.\, ,
\label{ih}
\end{align}
describes inelastic spin-flip processes in which magnons are created or destroyed as electrons traverse the tunnel barrier. These terms, $\mathcal{H}_e$ and $\mathcal{H}_i$, are field-theoretic representations of STT and MAT, respectively.

\section{charge transport}
\label{charge_trans}

In this section, we compute the charge current $I=(-e)\sum_{\nu,\sigma}\partial_t \langle \hat{b}_{\nu\sigma}^\dagger\hat{b}_{\nu\sigma}\rangle=-(-e)\sum_{\nu,\sigma}\partial_t \langle \hat{a}_{\nu\sigma}^\dagger\hat{a}_{\nu\sigma}\rangle$ into the right-lead perturbatively to second order in the amplitudes $A_{\nu\nu'}$ and $B_{\nu\nu'}$ (with $e>0$ as the negative of the electron charge), driven by an electrical bias across the MTJ. We take as our unperturbed Hamiltonian $\mathcal{H}_0=\sum_{\nu}\left(\epsilon_{\sigma \nu}  \hat{a}_{\nu\sigma}^\dagger\hat{a}_{\nu\sigma}+\epsilon_\nu  \hat{b}_{\nu\sigma}^\dagger\hat{b}_{\nu\sigma}\right)+\hbar \Omega \hat{c}^\dagger \hat{c}$, where $\epsilon_{\sigma \nu}$ and $\epsilon_{\nu }$ are the left and right-lead single particle energies, respectively, and $ \Omega $ is the ferromagnetic resonance frequency of the island spin. To simplify our model, we suppose the leads are good spin reservoirs, so that no spin accumulates there.

To second order in the tunneling amplitudes, we obtain a charge current for both the parallel ($ \langle \mathbf{S}\rangle$ oriented in the $ \mathbf{z}$ direction) and antiparallel ($ \langle \mathbf{S}\rangle$ oriented in the $- \mathbf{z}$ direction) configurations with elastic and inelastic contributions,
\begin{equation}
I=I_e+I_i\, ,
\label{ietot}
\end{equation}
respectively arising from $\mathcal{H}_e$ and $\mathcal{H}_i$. Here
\begin{align}
I_e= G V\,, \,\,\, I_i= (G_\phi/e) \Delta  \hbar \Omega S^{-1}\left[ N_{-}-\eta P_{L}\left(N_{+}-N \right)\right]\, ,
\label{ieparts}
\end{align}
where $V=V_L-V_R$ is the voltage bias. In the elastic term, the conductance $G$ depends on the magnon number $N$ and is given by $G=G_0+\eta G_m (1-N/S)$, with
\begin{align}
\nonumber
G_0 \equiv  G_\phi\sum_{\nu\nu'}\mathcal{A}^{(+)}_{L\nu' }(\epsilon_F)\mathcal{A}_{R\nu}(\epsilon_F)(|A_{\nu\nu'}|^2+|B_{\nu\nu'}|^2)\, ,\\
G_m \equiv 2 G_\phi\sum_{\nu\nu'}\mathcal{A}^{(-)}_{L\nu' }(\epsilon_F)\mathcal{A}_{R\nu}(\epsilon_F)\mathrm{Re}\left[A_{\nu\nu'}^*B_{\nu\nu'}\right]\, ,
\label{pols}
\end{align}
where we have assumed a flat electronic density of states near the Fermi surfaces $\epsilon_F$ of the leads. Here, $G_\phi=2e^2/h$ is the spin-degenerate quantum of conductance, and $\eta=\pm1$ for parallel (antiparallel) transport. Furthermore,  $\mathcal{A}_{R\nu }(\epsilon)$ is the right-lead spectral function, while $\mathcal{A}_{L\nu '}^{(\pm)}(\epsilon)=(\mathcal{A}_{L\uparrow \nu '}(\epsilon)\pm \mathcal{A}_{L\downarrow \nu' }(\epsilon))/2$, with $\mathcal{A}_{L\sigma \nu' }(\epsilon)$ as the spin-$\sigma$ left-lead spectral function.

In the inelastic term in Eq.~(\ref{ieparts}), $\Delta$ is a dimensionless parameter quantifying inelastic charge transport:
\begin{equation}
\Delta \equiv 2 \sum_{\nu\nu'}\mathcal{A}^{(+)}_{L\nu' }(\epsilon_F)\mathcal{A}_{R\nu}(\epsilon_F)| B_{\nu\nu'}|^2\, ,
\end{equation}
which vanishes when the island is nonmagetic, while $P_L$ is an effective left-lead polarization:
\begin{equation}
P_L\equiv 2 \sum_{\nu\nu'}\mathcal{A}^{(-)}_{L\nu' }(\epsilon_F)\mathcal{A}_{R\nu}(\epsilon_F)| B_{\nu\nu'}|^2/\Delta\, .
\end{equation}
When magnetism of the island is turned off ($B_{\nu\nu'}=0$), transport is elastic and the total charge current reduces to $I=G_0 V$. Last,
\begin{align}
\nonumber
N_{\pm}(V)\equiv\frac{1}{2\hbar \Omega}\left\{ n_{B}\left(\hbar \Omega
-eV\right) \left[\hbar \Omega-eV\right] \right. \\
\pm \left. n_{B}\left(\hbar \Omega+eV\right)\left[\hbar \Omega+eV\right] \right\}\, 
\label{Npm}
\end{align}
describes electron-hole excitations in the leads, with $n_B[\epsilon]=[e^{\epsilon/T}-1]^{-1}$ as a Bose-Einstein distribution, and $T$ the temperature in units of energy. The dependence of $I_i$ on $V$ through the functions $N_{\pm}(V)$ captures MAT, i.e. the alteration of charge transport caused by the absorption or emission of a magnon energy quantum $\hbar \Omega$ by an electron traversing the tunnel barrier.

The transport coefficients in Eq.~(\ref{ieparts}) allow for a more transparent parametrization in the simplifying case $A=A_{\nu\nu'}$ and $B=B_{\nu\nu'}$. There, $P_L$ reduces to traditional definition of polarization, $P_L=(D_{L\uparrow}-D_{L\downarrow})/(D_{L\uparrow}+D_{L\downarrow})$, where $D_{L\sigma}=\sum_{\nu'}\mathcal{A}_{L \sigma \nu'}(\epsilon_F)$.  Defining a complex island polarization $P\equiv B/A$, one finds that $G_0=G_\phi D_L D_R |A|^2 (1+|P|^2)$, $G_m=2 G_0 \mathrm{Re}[P] P_L/(1+|P|^2)$, and $\Delta =2 (G_0/G_\phi) |P|^2/(1+|P|^2)$, where $D_L=(D_{L\uparrow}+D_{L\downarrow})/2$ and $D_R=\sum_{\nu}\mathcal{A}_{R \sigma \nu}^{(+)}(\epsilon_F)$ are the left-lead spin-averaged and right-lead densities of states. 

 In the limit of infinite $S$, magnetic fluctuations of island spin do not contribute to the current. Here, the inelastic term in Eq.~(\ref{ietot}) vanishes, while the elastic term reduces to the classical expression for current traversing an MTJ with fixed orientations of the magnetic leads, $I= (G_0+\eta G_m) V$, i.e. a linear dependence of $I$ on $V$ that depends on the island orientation through $\eta$. 
 
 At finite $S$, however, the current depends nonlinearly on the bias $V$ through the electron-hole functions $N_{\pm}$ and the magnon number $N$. The dependence of $G_m$ on $N=S-\langle \hat{S}_z \rangle/\hbar$ can be interpreted as a change in the magnetoconductance $\delta G_m=\left(-N/S\right)G_m$ due to the average misorientation of the fluctuating island spin away from left-lead polarization direction $\mathbf{z}$. At zero bias, the current vanishes if $N=N_+(0)=n_B(\hbar\Omega)\equiv N_0$, i.e. the magnons are in equilibrium with the electron-hole excitations in the leads. At finite bias, however, $N$ is driven out-of-equilibrium and depends on $V$. To obtain the full dependence of $I$ on $V$, then, we turn to spin-transfer and magnon dynamics.

\section{Self-Consistent Spin Dynamics} 
\label{scsd}

In this section, we compute the bias dependence of the magnon occupation number in the steady-state, which is driven from equilibrium by the STT.  The same inelastic scattering processes described above drive magnetic dynamics. \cite{Kozub:2007gv} The corresponding island spin dynamics can be captured by a simple rate equation for the magnon number:
\begin{equation}
\hbar \dot{N}=-2\alpha_p \hbar \Omega \left(N-N_0 \right)+I_m\, ,
\end{equation}
where $I_m$ is the rate of angular momentum transfer to the island spin from the leads.  The damping coefficient $\alpha_p$ parametrizes coupling of the spin $\mathbf{S}$ to the lattice, which, in the absence of $I_m$, equilibrates $N$ to $N_0$. The spin current $I_m$ is obtained by calculating the rate of change of the $z$-component of angular momentum of the leads, $I_L^{(s)}=(\hbar/2)\sum_{\nu,\sigma\sigma'} \sigma^{z}_{\sigma\sigma'} \partial_t\langle \hat{a}_{\nu\sigma}^\dagger\hat{a}_{\nu\sigma'}\rangle$ and $I_R^{(s)}=(\hbar/2)\sum_{\nu,\sigma\sigma'}\sigma^{z}_{\sigma\sigma'} \partial_t \langle \hat{b}_{\nu\sigma}^\dagger\hat{b}_{\nu\sigma'}\rangle$, and exploiting conservation of total $z$-spin by $\mathcal{H}_T$, i.e. $I_m=-I_L^{(s)}-I_R^{(s)}$. One obtains again to second order:
\begin{equation}
I_m=-2\alpha_e \hbar \Omega \left(N-N_+ \right)+2\alpha_e P_L\left(N_ -\hbar \Omega+N V \right)\, ,
\label{im}
\end{equation}
where $\alpha_e=\Delta / S\pi$, so that electron fluctuations become increasingly important with decreasing junction size $\sim S$.  The first term, $\propto N-N_+=(1+N_+)N-N_+(1+N)$, is the difference between the rate of electron-hole emission/magnon absorption ($\propto 1+N_+$) and the rate of magnon emission/electron-hole absorption ($\propto 1+N$), and is nonzero due to the noncancellation of the spontaneous magnon and electron-hole emission terms. At zero bias, the second term in Eq.~(\ref{im}) vanishes, leaving $\hbar \dot{N}=-2(\alpha_e+\alpha_p)\hbar\Omega(N-N_0)$; here the rate of change of the magnon occupation number is the difference between the emission rate due to driving by fluctuations of phonons and electron-holes in the leads, and the absorption rate corresponding to dissipation of the magnetic dynamics back into the leads and lattice.

\begin{figure}[pt]
\includegraphics[width=\linewidth,clip=]{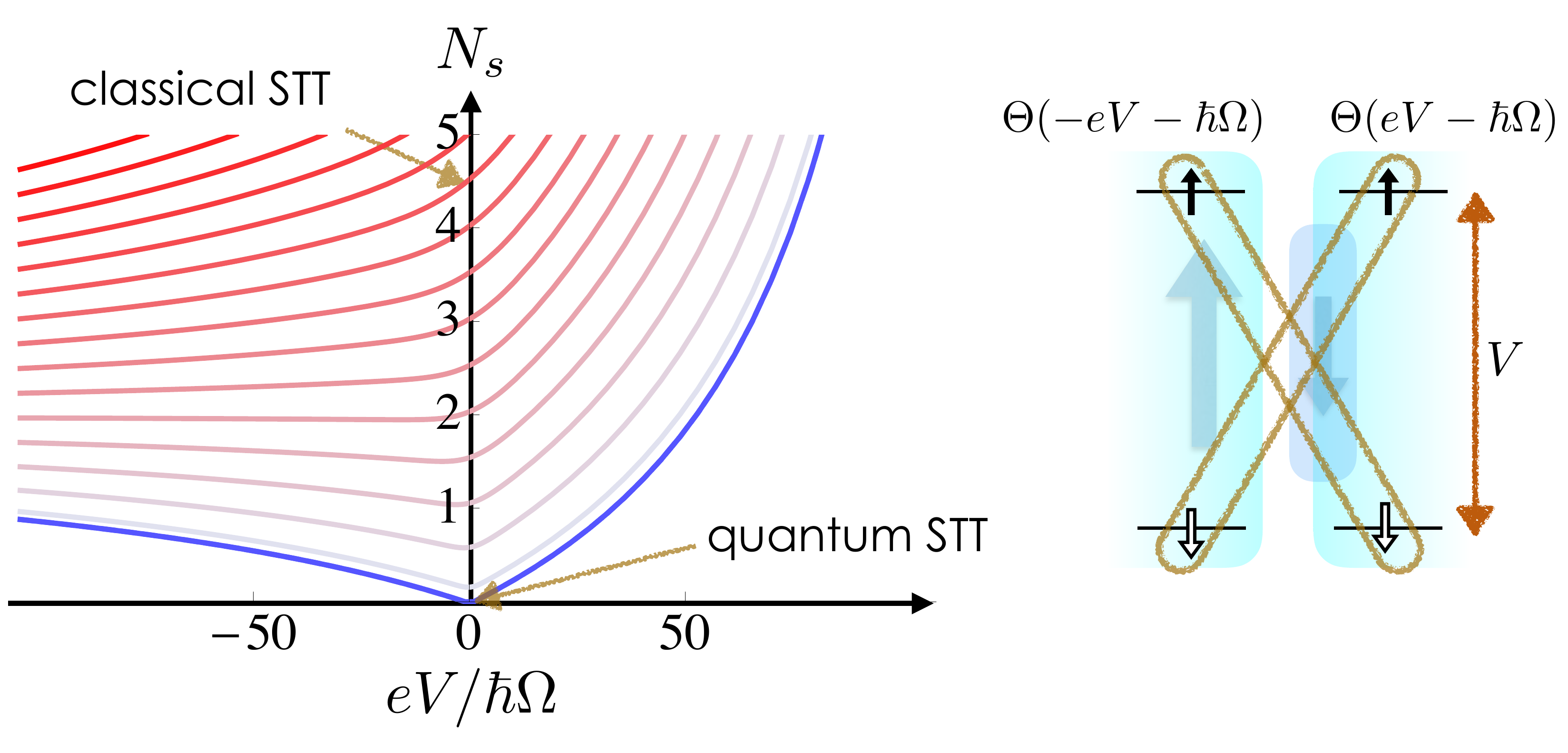}
\caption{Left figure: steady-state magnon number $N_s$, Eq.~(\ref{NsV}), versus bias $V$ for $T/\hbar \Omega=0$ (blue) to $10$ (red) in units of $1$, for the  antiparallel MTJ configuration with $\alpha_p/\alpha_e=25$. The symmetry between positive and negative $V$ is broken by the left-lead polarization $P_L=0.2$. Employing the parameterization from the discussion below Eq.~(\ref{Npm}) and taking $P=P_L$, one has $\Delta (G_\phi/G_0)=(G_m/G_0)=P^2/(1+P^2)$. At high-temperatures, the STT gives rise to a monotonically increasing magnon occupation number, while at low temperatures, a minimum near zero bias is formed. Upper right inset: electron/hole pairs giving rise to spin and charge transport at zero temperature. }.
\label{ns}
\end{figure} 

In the steady-state ($\dot{N}=0$), the out-of-equilibrium magnon number is given by $N=N_s(V)$, with
\begin{equation}
N_s(V)=\gamma N_0+\gamma \tilde{N}\, ,
\label{NsV}
\end{equation}
which is plotted in Fig.~\ref{ns}. Here $\gamma =1/(1+V/V_c)$ and $\tilde{N}=(\alpha_e/\alpha_p)(N_+(V)-N_0-P_L N_-(V))/(1+\alpha_e/\alpha_p)$. The quantity $V_c=\eta(1+\alpha_p/\alpha_e)\hbar \Omega /eP_L$ is the voltage threshold for the so-called ``swasing" instability;\cite{Berger:1996is} we restrict ourselves to biases $|V|<V_c$. The deviation of the magnon population away from the equilibrium value $N_0$ may interpreted as STT-induced alteration of the magnetic state of the island. The first term in Eq.~(\ref{NsV}) represents thermal fluctuations $N_0$ of $\mathbf{S}$, which are enhanced by a factor $\gamma$ by STT; when $P_L=0$, $V_c$ diverges, and this term no longer depends on $V$. The second term in Eq.~(\ref{NsV}), which vanishes at $V=0$, is an effective STT arising from inelastic electron-magnon scattering. This term survives at $P_L=0$, as impinging out-of-equilibrium electrons are known to depolarize spin even without a polarizing layer,\cite{Foros:2005fy} acting as an effective heating of the ferromagnetic island spin.\cite{Nunez:2008cz} In the limit of infinite $S$, one has $\alpha_e$ and therefore $V_c$ diverge, so $N_s(V)$ is fixed at $N_0$, and $\mathbf{S}$ is no longer altered by the STT. 

According to Eq.~(\ref{NsV}), two types of temperature regimes for STT may be distinguished.  At high temperatures where $N_0 \gg \tilde{N}$ so $N\approx \gamma N_0$, classical STT dominates, and $N$ depends $monotonically$ on $V/V_c$. Here, the bias dependence of $N$ thus arises from the STT-alteration of $thermal$ fluctuations of $\mathbf{S}$.   At low temperatures where $N_0\ll \tilde{N}$, the classical interpretation of STT-enhanced thermal fluctuations of $\mathbf{S}$ no longer holds, and quantum STT dominates. Since $\tilde{N}/N_0\sim (\alpha_e/\alpha_p)/(1+\alpha_e/\alpha_p)\sim 1/(const+S)$, the classical-to-quantum transition occurs at lower temperatures for larger $S$, suggesting that such a quantum effect will manifest only in sufficiently small junctions or sufficiently low temperatures.  Even as $T\rightarrow 0$, where $N_0=0$,  the junction exhibits a nonlinear electrical response. This zero temperature, quantum STT can be understood as follows. Thermal fluctuations of $\mathbf{S}$ and electron-hole pairs within each lead freeze out.  However, electron-hole pairs $across$ the junction, split by $V$, are available for inelastic magnon scattering (see schematic, right side of Fig.~\ref{ns}): because $n_{B}\left(\hbar \Omega\pm eV\right)=-\Theta \left( \mp e V-\hbar \Omega \right)$ at zero temperature, where $\Theta(x)$ is the step function, the electron-hole functions $N_{\pm}$ are nonzero when $e\left| V \right| >\hbar \Omega$. These pairs drive a nonequilibrium magnon population $\gamma \tilde{N}$, which can be interpreted as the generation of an effective nonzero magnon temperature.\cite{Nunez:2008cz}

\section{Electrical response of MTJ}
\label{elec_respons}

\begin{figure}[pt]
\includegraphics[width=0.8\linewidth,clip=]{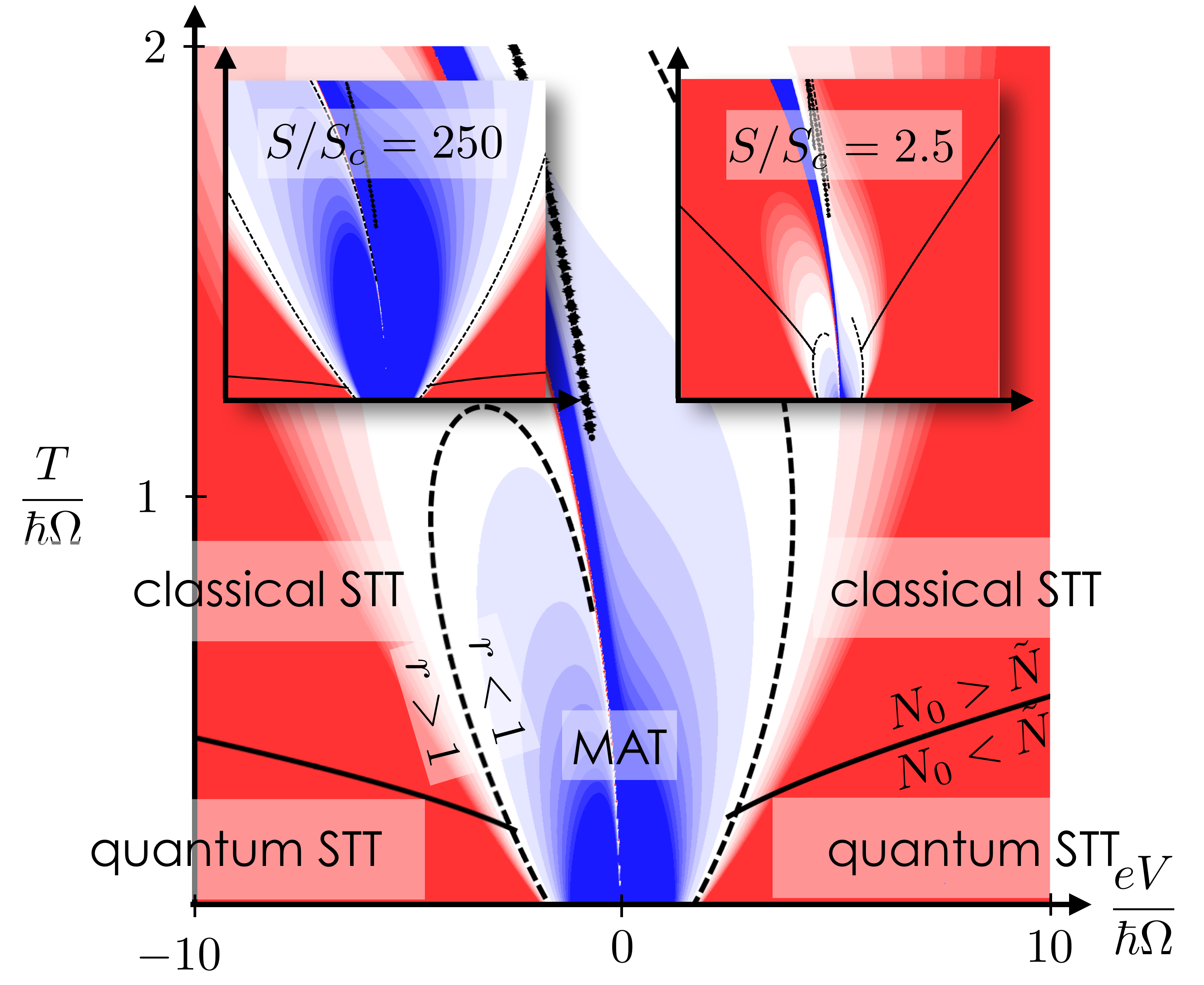}
\caption{Plot of the ratio of $r=\mathcal{G}'_{\mathrm{STT}}/\mathcal{G}'_{\mathrm{MAT}}$ as a function of bias $V$ and temperature $T$, demonstrating the relative importance of STT to MAT in the nonlinear charge transport features. MAT dominates over STT at low temperatures and biases.  Blue (red) contours correspond MAT (STT) dominating, i.e. to $r<1$($r>1$) in steps of $1/2$. The black, solid lines correspond to $N_0=\tilde{N}$; at temperatures above this line, $N_0>\tilde{N}$ and classical STT-driven thermal fluctuations dominate the response $\mathcal{G}_{\mathrm{STT}}$; below,  $\tilde{N}>N_0$,  and inelastic-scattering dominates. The parameters are the same as Fig.~\ref{ns}, so $S/S_c=25$. Upper left and right insets: $r$ for $S/S_c=250$ and $S/S_c=2.5$ respectively, with the same axes-scale as the main figure. For $S<S_c$, $r$ no longer depends on $S$; for $S\gg S_c$, MAT dominates over STT for a larger range of temperatures and biases, reflecting the decreased importance of fluctuations of $\mathbf{S}$.}
\label{nlin}
\end{figure}

\begin{figure}[pt]
\includegraphics[width=\linewidth,clip=]{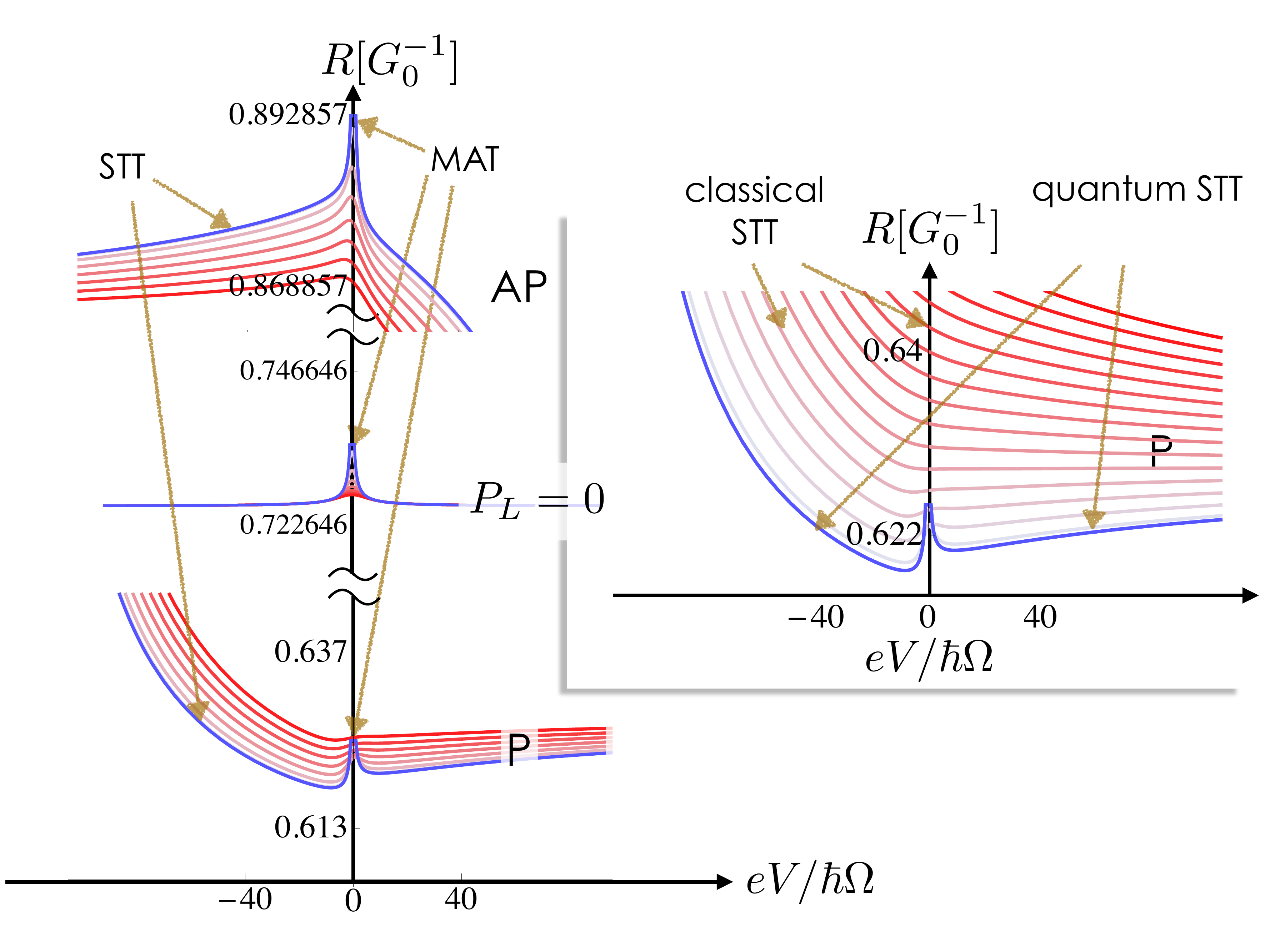}
\caption{Main figure: low temperature resistance $R$ versus bias for parallel (P) and antiparallel (AP) configurations, respectively corresponding to $\eta=\pm1$, as well as $P_L=G_m=0$ (left-lead unpolarized). Shown are temperatures ranging from $T/\hbar\Omega=0$ (blue) to $2$ (red) in steps of $1/4$. MAT dominates at temperatures $T\ll \hbar \Omega$ and $e|V| \lesssim \hbar \Omega$, but vanishes as the temperature is raised. Inset: high temperature $R$ versus bias for the parallel configuration, for temperatures ranging from $T/\hbar\Omega=0$ (blue) to $8$ (red) in steps of $1/2$. At $T=0$, the resistance exhibits a maximum around zero bias, reflecting MAT. At $T/\hbar \Omega\sim1$ the resistance exhibits a minimum near zero bias, stemming from quantum STT. As $T$ is further increased, classical STT dominates, and $R$ no longer shows a local extrema near zero bias. All unspecified parameters are the same as in Fig.~\ref{ns}.  We find that the MAT-induced resistance plateau survives at $P_L=G_m=0$, as only one magnetic component is required to inelastically scatter electrons. } 
\label{resis}
\end{figure}

Let us summarize our key results: Eqs.~(\ref{ietot}),~(\ref{ieparts}) and~(\ref{NsV}). In order to obtain the full dependence of the current on voltage, we insert $N=N_s(V)$ into Eq.~(\ref{ietot}). Two effects give rise to a nonlinear relationship between $I$ and $V$: MAT (dependence of $I_i$ on $V$ via $N_\pm(V)$) and STT (dependence of $I$ on $V$ via the magnon distribution $N$). In an experiment, these effects may manifest as nonlinear features in, e.g., the electrical resistance $R=V/I$. In order to parse which effect dominates the nonlinearity in $R$ at a given temperature and bias, it is helpful to write the differential conductance $\mathcal{G}=dI/dV$ as:
\begin{equation}
\mathcal{G}=G_0+\eta G_m+\mathcal{G}_{\mathrm{MAT}}+\mathcal{G}_{\mathrm{STT}}\, ,
\end{equation}
where $\mathcal{G}_{\mathrm{MAT}}\equiv \left. \partial_V I_i\right|_{N=N_s(V)}$ is the differential conductance arising from MAT and $\mathcal{G}_{\mathrm{STT}}\equiv \left. \partial_N I \right|_{N=N_s(V)} \partial_V N_s(V) $ is the differential conductance STT. To isolate the \textit{nonlinear} features, we consider the ratio $r\equiv \partial_V\mathcal{G}_{\mathrm{STT}}/\partial_V\mathcal{G}_{\mathrm{MAT}}$ \footnote{The quantity  $\partial_V\mathcal{G}_{\mathrm{MAT}}$  involves a term $\propto \partial_V N$, corresponding to an interplay of both STT and MAT; however, for small polarizations, $P_L\ll1$, this term is subdominant to other terms.}, which is plotted in Fig.~\ref{nlin}. The resistance $R$ at various temperatures is shown in Fig.~\ref{resis}.

At low temperatures ($T\ll \hbar \Omega$) and biases ($e|V|\lesssim  \hbar\Omega$), $r \ll 1$ (blue regions of Fig.~\ref{nlin}), and MAT dominates the dependence of the resistance $R=V/I$ on $V$.  Here we find that for all orientations of the MTJ, i.e. both signs of $\eta$, MAT manifests as a \textit{plateau} in the resistance\cite{Zhang:1997kx,Han:2001er,Lu:2003hv,Balashov:2008ih} (see Fig.~\ref{resis}).  Note that generally features of MAT due to macrospin fluctuations manifest on the bias scale $e|V|\sim \hbar \Omega$. 

At high temperatures ($T \gtrsim \hbar \Omega$) and/or biases ($e|V| \gg \hbar \Omega$), $r \gg 1$ (red regions of Fig.~\ref{nlin}), and STT determines the nonlinear of behavior of $R$. The two STT temperature regimes, corresponding to classical STT ($N_0 \gg \tilde{N}$) and quantum STT ($\tilde{N} \gg N_0$), as discussed above, give rise to different behaviors of the resistance. At higher temperatures where $N\approx \gamma N_0$, the resistance depends $monotonically$ on $V$ through $\gamma$ near zero bias; at lower temperatures where $N\approx \gamma \tilde{N}$, the resistance, like $N$, shows an extremum (see inset, Fig.~\ref{resis}). The transition from a monotonic dependence of $R$ on $V$ reflects the change from classical spin fluctuations $\gamma N_0$, which are enhanced only for one direction of current, to $\gamma \tilde{N}$, which are enhanced for $either$ direction. Such a transition, from monotonically changing $R$ to extremum, is seen clearly in [\onlinecite{Zholud:2017ed}]. For the parameters chosen in Fig.~\ref{resis}, the classical-to-quantum occurs at a temperature near $T\sim\hbar \Omega$. Importantly, unlike the effect of MAT, which is observable in the range $e|V|\sim \hbar \Omega$, the STT extrema persist over an energy range $eV_c$ ($\gg \hbar \Omega$ for $P_L\ll 1$), due to $\gamma$. As with MAT, these features survive at zero-temperature.

It should be noted that the value of $S$, which can be assumed to scale with the junction size, also determines the regions in which classical STT, quantum STT and MAT dominate the nonlinear signal (see upper insets of Fig.~\ref{nlin}). The quantity $S_c\equiv \Delta/\pi \alpha_p$, defined so that $S/S_c =\alpha_p/\alpha_e$, provides a convenient reference value. For larger values of $S/S_c$, STT-driving of $N$ away from $N_0$ is increasingly suppressed, and MAT dominates over a wider range of biases and temperatures, with our results reducing to those of [\onlinecite{Zhang:1997kx}] in those ranges. In addition, when $|V|$ is sufficiently large that STT dominates over MAT, the temperatures below which quantum STT dominates over classical STT (corresponding to $\tilde{N}<N_0$), decreases with increasing values of $S/S_c$, reflecting the suppression of quantum spin fluctuations. This helps to explain why quantum STT may be expected to play a significant role in the electrical properties of small (in the case of [\onlinecite{Zholud:2017ed}], \textit{nano}scale) MTJs. 

The various parameters are readily estimated. Typical ferromagnetic resonance frequencies $\Omega \sim 10 \textrm{ GHz}$ $\sim 10^{-6} \textrm{eV}/\hbar$ corresponds to a crossover temperature of about $0.1 \textrm{ K}$. For external field strengths of $\sim 1\,T$, however, comparable to those used in [\onlinecite{Zholud:2017ed}] and corresponding to resonance frequencies $\sim 100\textrm{ GHz}\sim 10^{-5} \textrm{eV}/\hbar$, the crossover temperature becomes $\sim 1 \textrm{ K}$, which can be further increased by decreasing $S$ (see upper right inset of Fig.~\ref{nlin}), decreasing $\alpha_p$ (thereby increasing $S_c$), or increasing the bias. Taking $G_0 \sim \Omega^{-1}$ and $P\sim 10^{-1}$, corresponds to $\Delta \sim 10^{6}$; for a conservative value $\alpha_p=10$, this corresponds to a reference spin $S_c \sim 10^5$.

We conclude this section with qualitative predictions for experiments. First, as both quantum STT and MAT depend on the functions $N_{\pm}(V)$, at low temperatures $T\ll \hbar \Omega$ the resistance is flat at biases below the magnon gap, $e|V| \leq \hbar \Omega$, as impinging electrons are not sufficiently energetic to excite magnons. Thus, we expect that at sufficiently large magnetic fields, the low-temperature resistance for biases $e|V|\leq \hbar \Omega$ should be flat for nanoscale junctions wherein the micromagnetic modes are gapped out and our macrospin model is valid. For the parameters in [\onlinecite{Zholud:2017ed}], for example, this corresponds to a current range of $|I|<0.1$mA, which may be beyond the resolution of the experiment.  Second, at low temperatures, one may determine whether MAT or STT dominates by changing the orientation of the magnetic junction. The zero-bias extrema in the resistance due to MAT are always plateaus in the resistance, whereas those due to quantum STT are valleys (plateaus) in the parallel (antiparallel) configuration\cite{Sun:2008ik,Fuchs:tb,Sankey:2007fg,Manchon:2008we,Zholud:2017ed} (see Fig.~\ref{resis}).  This is to be expected, as inelastic scattering of electrons by magnons does not require a polarizer (scattering by phonons, for example, will show similar behavior), while the magnetoconductance clearly depends on the sign of $\eta$.

\section{Conclusion and Discussion}
\label{disc}

We have demonstrated how a combination of inelastic charge transport and STT generate a rich dependence of the resistance on voltage. In particular, we have shown how MAT and STT amplification of both equilibrium and nonequilibrium spin fluctuations driven by inelastic scattering can generate a low temperature nonlinear resistance similar to that observed experimentally \cite{Sun:2008ik,Zholud:2017ed}.

Future work may expand on our model. For simplicity, this model treats only one magnon mode, but in general, at high temperatures a spectrum of modes may contribute, enhancing the nonlinear features; if the spectrum is known, our model can be adapted accordingly by weighing the current, Eq.~(\ref{ietot}), by a density of states and integrating over magnon energy $\hbar \Omega$. The incorporation of higher energy ($>\hbar \Omega$), micromagnetic excitations extends the temperatures at which MAT can be studied.\cite{Zhang:1997kx} Joule heating, absent in our model, can generate thermal fluctuations of the spin, even if the ambient temperature is low; while such an effect is dismissed in [\onlinecite{Zholud:2017ed}], it must be generally addressed. Furthermore, our theory is perturbative in the tunneling coefficients $A_{ \nu \nu'}$ and  $B_{ \nu \nu'}$, and we neglect higher order terms in the tunneling coefficients,\cite{Mahfouzi:2014hu} which for small values lead to Kondo correlations at low temperatures. Absent in our model are also nonparabolic electronic band structure features of the normal metals, which can give rise to a nonlinear resistance. In the case when the island is physically separate from right lead, mesoscopic low-bias anomalies in spin-dependent transport \cite{Useinov:2017eg,*Esmaeili:er} may result, which could obscure the magnetotransport features studied in this paper.

  In addition, when $V$ approaches the swasing threshold, $N$ can become on the order of $S$, so that our expansion of the radical in Eq.~(\ref{hp}) breaks down, and a more careful treatment is required.  While classical current-driven instabilities in MTJs have already been observed,\cite{Huai:2004bd,Ralph:2008kj} it remains to be seen how fluctuations of the magnetic order modify charge transport for $V$ greater than $V_c$.

S.A.B. and R.D. are supported by funding from the Stichting voor Fundamenteel Onderzoek der Materie (FOM) and the European Research Council via Advanced Grant number 669442 ``Insulatronics''. Y.T. is supported by NSF under Grant No. DMR-1742928.


%

\end{document}